\documentclass[11pt]{article}
\usepackage{amsmath, amssymb, amstext}
\def\mxth{\mathsurround=0pt }
\def\xversim#1#2{\lower2.pt\vbox{\baselineskip0pt \lineskip-.5pt
\ialign{$\mxth#1\hfil##\hfil$\crcr#2\crcr\sim\crcr}}}

\begin{document}
\begin{titlepage}
\vskip 8mm
\begin{flushright}
YITP-05-12 \\
ITP-UU-05/12 \\
SPIN-05/10
\end{flushright}
\vskip 20mm
\begin{center}
  {\Large{\bf Introduction to black hole entropy and supersymmetry
    }}\footnote[1]{Based on lectures presented at the III Summer 
    School in Modern Mathematical Physics, Zlatibor, 20 -- 31 August, 
    2004.}
\end{center}
\vskip 12mm

\begin{center}
{\bf Bernard de Wit}\\[3mm]
Yukawa Institute, Kyoto University, Kyoto, Japan  \\[1mm] 
Institute for Theoretical Physics \,\&\, Spinoza Institute,\\  
Utrecht University, The Netherlands\\[1mm]
{\tt b.dewit@phys.uu.nl} 
\end{center}

\vskip .6in

\begin{center} {\bf Abstract } \end{center}
\begin{quotation}\noindent
  In these lectures we introduce some of the principles and techniques
  that are relevant for the determination of the entropy of extremal
  black holes by either string theory or supergravity. We
  consider such black holes with $N=2$ and $N=4$ supersymmetry,
  explaining how agreement is obtained for both the terms that are
  leading and those that are subleading in the limit of large charges.
  We discuss the
  relevance of these results in the context of the more recent developments.
\end{quotation}
\end{titlepage}
\eject
\section{Introduction}
The aim of these lectures is to present a pedagogical introduction to
recent developments in string theory and supergravity with regard to
black hole entropy. In the last decade there have been many advances
which have thoroughly changed our thinking about black hole entropy.
Supersymmetry has been an indispensable ingredient in all of this, not
just because it is an integral part of the theories we consider, but
also because it serves as a tool to keep the calculations tractable.
Comparisons between the macroscopic and the microscopic entropy, where
the latter is defined as the logarithm of the degeneracy of
microstates of a certain brane or string configuration, were first
carried out in the limit of large charges, assuming the
Bekenstein-Hawking area law on the macroscopic (supergravity) side.
Later on also subleading corrections could be evaluated on both sides
and were shown to be in agreement, although the area law ceases to
hold.  More recently these results have rekindled the interest in
these questions and during the past year a variety of new developments
took place.

We start with a brief exposition about the relation between black hole
mechanics and thermodynamics. Then we review the connection between
microscopic and macroscopic descriptions of black holes. As an example
we discuss the situation for Calabi-Yau black holes, both from a
microscopic and a macroscopic perspective. We elucidate the presence
of subleading corrections, which, on the supergravity side originate
from interactions with higher-order derivative terms.  Subsequently we
discuss heterotic black holes, where we extend the discussion to $N=4$
supersymmetry introducing non-holomorphic corrections in order to
preserve $S$-duality invariance. These results for $N=4$ dyonic black
holes are successfully confronted with a microscopic degeneracy
formula. After a brief discussion of the $N=4$ purely electric black
holes and their relation to perturbative string states, we introduce
the so-called mixed black hole partition function and discuss some of
its consequences.

\section{Black holes and thermodynamics}
\label{bh+th}
Black holes are, roughly speaking, solutions of Einstein's equations
of general relativity that exhibit an event horizon. From inside this
horizon, nothing (and in particular, no light) can escape. In the
context of this talk, it suffices to think of spherically symmetric
and static black holes, with a flat space-time geometry at spatial
infinity. By definition, the region inside the horizon is not in the
backward lightcone of future timelike infinity. However, since the
discovery of Hawking radiation it has become clear that many of the
above classical features of black holes will have to be modified.

We will be considering black holes in four space-time dimensions,
carrying electric and/or magnetic charges. Such solutions can be
described by Einstein-Maxwell theory, the classical field theory for
gravity and electromagnetism. The most general static black holes of
this type correspond to the Reissner-Nordstrom solutions. They are
characterized by a charge $Q$ and a mass $M$. In the presence of
magnetic charges, $Q$ is replaced by $\sqrt{q^2+p^2}$ in most
formulae, where $q$ and $p$ denote the electric and the magnetic
charge, respectively. In this section there will therefore be no need
to distinguish between the two types of charges. Note that without
charges we will just be dealing with Schwarzschild black holes.

Two quantities associated with the black hole horizon are the area $A$
and the surface gravity $\kappa_{\rm s}$. The area is simply the area
of the two-sphere defined by the horizon. The surface gravity, which
is constant on the horizon, is related to the force (measured at
spatial infinity) that holds a unit test mass in place. The mass $M$
and charge $Q$ of the black hole are not directly associated with the
horizon and can be expressed by appropriate surface integrals at
spatial infinity.

As is well known, there exists a striking correspondence between the laws of
thermodynamics and the laws of black hole mechanics \cite{BCH}. Of
particular importance is the first law, which, for thermodynamics,
states that the variation of the total energy is equal to the temperature
times the variation of 
the entropy, modulo work terms, for instance proportional to a change
of the volume. The corresponding formula for
black holes expresses how the variation of the black hole mass is
related to the variation of the horizon area, up to work terms
proportional to the variation of the angular momentum. In addition
there can also be a term proportional to a 
variation of the charge, multiplied by the electric/magnetic
potential $\phi$ at the horizon. Specifically, the first law of
thermodynamics, $\delta E =  
T\,\delta S - p\,\delta V$, translates into 
\begin{equation}
\label{first-law} 
\delta M = \frac{\kappa_{\rm s}}{2\pi} \, \frac {\delta A}{4}
+\phi\,\delta Q +\Omega\,\delta J\,. 
\end{equation} 
The reason for factorizing the first term on the right-hand side in
this particular form, is
that $\kappa_{\rm s}/2\pi$ represents precisely the Hawking
temperature \cite{Hawking}. This then leads to the identification of
the black hole entropy in terms of the horizon area, 
\begin{equation}
\label{area-law}
{\cal S}_{\rm macro} = \tfrac14 A\,,
\end{equation}
a result that is known as the area law \cite{Bek}.
In these equations the various quantities have been defined in Planck
units, meaning that they have 
been made dimensionless by multiplication with an appropriate power of
Newton's constant (we will set $\hbar=c=1$). This constant appears in the
Einstein-Hilbert  
Lagrangian according to ${\cal L}_{\rm EH} = -(16\pi\,G_{\rm N})^{-1}\,
\sqrt{\vert g\vert} \,R$. With this normalization the quantities
appearing in the first law are independent of the scale of the
metric. 

Einstein-Maxwell theory can be naturally embedded
into $N=2$ supergravity which may lead to an extension with a 
variety of abelian gauge fields and a related number of massless 
scalar fields (often called `moduli' fields, for reasons that will 
become clear later on). At spatial infinity these moduli fields will
tend to a constant, and the black hole mass will depend on these
constants, thus introducing additional terms on the right-hand side of
(\ref{first-law}).  

For Schwarzschild black holes the only relevant parameter is the mass
$M$ and we note the following relations, 
\begin{equation}
\label{schwarzschild}
 A= 16\pi\,M^2\,,\qquad \kappa_{\rm s}= \frac{1}{4\,M}\;,
\end{equation}
consistent with (\ref{first-law}). For the 
Reissner-Nordstrom black hole, the situation is more subtle. Here one
distinguishes three different cases. For $M>Q$ one has the
non-extremal solutions, which exhibit two horizons, an exterior event
horizon and an interior 
horizon. When $M=Q$ one is dealing with an extremal black hole, for
which the two horizons coincide and the surface gravity vanishes. In
that case one has
\begin{equation}
\label{e-RN}
 A= 4\pi\,M^2\,,\qquad \kappa_{\rm s}=0\,,\qquad \phi = Q\,
 \sqrt{\frac{4\pi}A}\,.  
\end{equation}
It is straightforward to verify that this result is consistent with
(\ref{first-law}) for variations in the subspace of extremal black
holes ({\it i.e.}, with $\delta M=\delta Q$). Because the surface gravity
vanishes, one might expect the entropy to vanish as 
suggested by the third law of thermodynamics. Obviously, that is not
the case as the horizon area remains finite for zero surface gravity.
Finally, solutions with $M<Q$ are not regarded as physically
acceptable. Their total energy is less than the electromagnetic
energy alone and they no longer have an event horizon but exhibit a
naked singularity. Hence 
extremal black holes saturate the bound $M\geq Q$ for physically
acceptable black hole solutions. 

When embedding Einstein-Maxwell theory into a complete supergravity
theory, the above classification has an interpretation in terms of the
supersymmetry algebra. This algebra has a central extension
proportional to the black hole charge(s). Unitary representations of
the supersymmetry algebra must necessarily have masses that are larger
than or equal to the charge. When this bound is saturated, one is
dealing with so-called BPS supermultiplets. Such supermultiplets are
smaller than the generic massive $N=2$ supermultiplets and have a different
spin content. Because of this, BPS states are stable under (adiabatic)
changes of the coupling constants, and the relation between charge
and mass remains preserved. This important feature of BPS states will
be relevant for what follows.

\section{On macroscopic and microscopic descriptions}
\label{sec:macro-micro}
A central question in black hole physics concerns the
statistical interpretation of the black hole entropy. String theory
has provided new insights here \cite{Strominger:1996sh}, which have
led to important results. In this context it is relevant that strings
live in more that four space-time dimensions. In most situations the
extra dimensions are compactified on some internal manifold $X$ and
one is dealing with the usual Kaluza-Klein scenario leading to
effective field theories in four dimensions, describing low-mass modes
of the fields associated with certain eigenfunctions on the internal
manifold.

Hence the original space-time will locally be a product $M^4\times X$,
where $M^4$ denotes the four-dimensional space-time that we experience
in daily life. We will denote the coordinates of $M^4$ by $x^\mu$ and
those of $X$ by $y^m$. In the situation described above there exists a
corresponding space $X$ at every point $x^\mu$ of $M^4$, whose size is
such that it will not be directly observable. However, this space $X$
does not have to be the same at every point in $M^4$, and moving
through $M^4$ one may encounter various spaces $X$ which may or may
not be equivalent.  Usually these spaces belong to some well-defined
class of fixed topology parametrized by certain moduli. These moduli
will appear as fields in the four-dimensional effective field theory.
For instance, suppose that the spaces $X$ are $n$-dimensional tori
$T^n$. The metric of $T^n$ will appear as a field in the
four-dimensional theory and is related to the torus moduli. Hence,
when dealing with a solution of the four-dimensional theory that is
not constant in $M^4$, each patch in $M^4$ has a nontrivial image in
the space of moduli that parametrize the internal spaces $X$.

Let us return to a black hole solution, viewed in this
higher-dimensional perspective. Now the fields, and in particular the
four-dimensional space-time metric, will vary nontrivially over $M^4$,
and so will the internal space $X$.  When moving to the center of the
black hole the gravitational fields will become strong and the
local product structure into $M^4\times X$ could break down.
Conventional Kaluza-Klein theory does not have much to say about what
happens, beyond the fact that the four-dimensional solution can be
lifted to the higher-dimensional one, at least in principle.

However, there is a feature of string theory that is absent in a
purely field-theoretic approach. In the effective field-theoretic
context only the local degrees of freedom of strings and branes are
captured. But extended objects may also carry global degrees of
freedom, as they can also wrap themselves around nontrivial cycles of
the internal space $X$. This wrapping tends to take place at a
particular position in $M^4$, so in the context of the four-dimensional
effective field theory this will reflect itself as a pointlike
object. This wrapped object is the string theory representation of the
black hole!

We are thus dealing with two complementary pictures of the black hole.
One based on general relativity where a point mass generates a global
solution of space-time with strongly varying gravitational fields,
which we shall refer to as the {\it macroscopic} description.  The
other one, based on the internal space where an extended object is
entangled in one of its cycles, does not immediately involve
gravitational fields and can easily be described in flat space-time.
This description will be refered to as {\it microscopic}. To
understand how these two descriptions are related is far from easy,
but a connection must exist in view of the fact that gravitons are
closed string states which interact with the wrapped branes. These
interactions are governed by the string coupling constant $g_{\rm s}$
and we are thus confronted with an interpolation in that coupling
constant. In principle, such an interpolation is very difficult to
carry out, so that a realistic comparison between microscopic and
macroscopic results is usually impossible. However, reliable
predictions are possible for extremal black holes! In a supersymmetric
setting extremal black holes are BPS and, as we indicated earlier, in
that situation there are reasons to trust such interpolations. Indeed,
it has be shown that the predictions based on these two alternative
descriptions can be successfully compared and new insights about black
holes can be obtained.

But how do the wrapped strings and branes represent themselves in the
effective action description and what governs their interactions with
the low-mass fields? Here it is important to realize that the massless
four-dimensional fields are associated with harmonic forms on $X$.
Harmonic forms are in one-to-one correspondence with so-called
cohomology groups consisting of equivalence classes of forms that are
closed but not exact. The number of independent harmonic forms of a
given degree is given by the so-called Betti numbers, which are fixed
by the topology of the spaces $X$. When expanding fields in a
Kaluza-Klein scenario, the number of corresponding massless fields can
be deduced from an expansion in terms of tensors on $X$
corresponding to the various harmonic forms. The higher-dimensional
fields $\Phi(x,y)$ thus decompose into the massless fields $\phi^A(x)$
according to (schematically),
\begin{equation}
\label{KK-decomposition}
\Phi(x,y) = \phi^A(x)\;\omega_A(y)\,,
\end{equation}
where $\omega_A(y)$ denotes the independent harmonic forms on $X$. The above
expression, when substituted into the action of the higher-dimensional
theory, lead to interactions of the fields $\phi^A$ proportional to
the `coupling constants', 
\begin{equation}
\label{interaction}
C_{ABC\cdots} \propto \int_X \omega_A\wedge\omega_B\wedge\omega_C
\cdots\,. 
\end{equation}
These constants are known as intersection numbers, for reasons that
will become clear shortly. 

We already mentioned that the Betti numbers depend on the topology of
$X$. This is related to Poincar\'e duality, according to which
cohomology classes are related to homology classes. The latter consist
of submanifolds of $X$ without boundary that are themselves not a
boundary of some other submanifold of $X$. This is precisely relevant
for wrapped branes which indeed cover submanifolds of $X$, but are not
themselves the boundary of a submanifold because otherwise the brane
could collapse to a point. Without going into detail, this implies
that there exists a dual relationship between harmonic $p$-forms
$\omega$ and $(d_X-p)$-cycles, where $d_X$ denotes the dimension of
$X$. We can therefore choose a homology basis for the $(d_X-p)$-cycles
dual to the basis adopted for the $p$-forms.  Denoting this basis by
$\Omega_A$ the wrapping of an extended object can now be characterized
by specifying its corresponding cycle ${\cal P}$ in terms of the homology
basis,
\begin{equation}
\label{intersection}
{\cal P} = p^A\,\Omega_A\,.
\end{equation}
The integers $p^A$ count how many times the extended object is wrapped
around the corresponding cycle, so we are actually dealing with
integer-valued cohomology and homology. The wrapping numbers $p^A$
reflect themselves as magnetic charges in the effective action. The
electric charges are already an integer part of the effective action,
because they are associated with gauge transformations that usually
originate from the higher-dimensional theory. 

Owing to Poincar\'e duality it is thus very natural that the winding
numbers interact with the massless modes in the form of magnetic
charges, so that they can be incorporated in the effective action.
Before closing this section, we note that, by Poincar\'e duality, we
can express the number of intersections by
\begin{equation}
\label{intersections} 
P\cdot P\cdot P\cdots =C_{ABC\cdots}\,p^Ap^Bp^C\cdots\;.
\end{equation}
This is a topological characterization of the wrapping, which will
appear in later formulae.

\section{Black holes in M/String Theory}
\label{sec:M-bh} 
As an example we now discuss the black hole entropy derived from both
microscopic and macroscopic arguments in a special case.  We start
from M-theory, which, in the strong coupling limit of type-IIA string
theory, is described by eleven-dimensional supergravity. The latter is
invariant under 32 supersymmetries. Seven of the eleven space-time
dimensions are compactified on an internal space which is the product
of a Calabi-Yau threefold (a three-dimensional complex manifold, which
henceforth we denote by $CY_3$) times a circle $S^1$. Such a space
breaks part of the supersymmetries and only 8 of them are left
unaffected. In the context of the four-dimensional space-time $M^4$,
these 8 supersymmetries are encoded into two independent Lorentz
spinors and for that reason this symmetry is referred to as $N=2$
supersymmetry. Hence the effective four-dimensional field theory will
be some version of $N=2$ supergravity.

M-theory contains a five-brane and this is the microscopic object that is
responsible for the black holes that we consider; the five-brane has
wrapped itself on a 4-cycle ${\cal P}$ of the ${CY}_3$ space
\cite{Maldacena:1997de}.  Alternatively one may consider this class of
black holes in type-IIA string theory, with a 4-brane wrapping the
4-cycle \cite{Vafa:1997gr}. The 4-cycle is subject to certain
requirements which will be mentioned in due course.

The massless modes captured by the effective field theory correspond
to harmonic forms on the $CY_3$ space; they do not depend on the $S^1$
coordinate. The 2-forms are of particular interest. In the effective
theory they give rise to vector gauge fields $A_\mu{}^A$, which
originate from the rank-three tensor gauge field in eleven dimensions.
In addition there is an extra vector field $A_\mu{}^0$ coresponding to
a 0-form which is related to the graviphoton associated with $S^1$.
This field will couple to the electric charge $q_0$ associated with
momentum modes on $S^1$ in the standard Kaluza-Klein fashion. The
2-forms are dual to 4-cyles and the wrapping of the five-brane is
encoded in terms of the wrapping numbers $p^A$, which appear in the
effective field theory as magnetic charges which couple to the gauge
fields $A_\mu{}^A$. Here we see Poincar\'e duality at work, as the
magnetic charges couple nicely to the corresponding gauge fields. For
a Calabi-Yau three-fold, there is a triple intersection number
$C_{ABC}$, which appears in the three-point couplings of the effective
field theory. There is a subtle topological feature that we have not
explained before, which is typical for complex manifolds containing 
4-cycles, namely the existence of another quantity of topological
interest known as the second Chern class. The second Chern class is a
4-form whose integral over a four-dimensional Euclidean space defines
the instanton number. The 4-form can be integrated over the 4-cycle
${\cal P}$ and yields $c_{2A} \,p^A$, where the $c_{2A}$ are integers.

Let us now turn to the microscopic counting of degrees of freedom
\cite{Maldacena:1997de}.  These degrees of freedom are associated with
the massless excitations of the wrapped five-brane characterized by
the wrapping numbers $p^A$ on the 4-cycle. The 4-cycle ${\cal P}$ must
correspond to a holomorphically embedded complex submanifold in order
to preserve 4 supersymmetries.  The massless excitations of the
five-brane are then described by a $(1+1)$-dimensional superconformal
field theory (the reader may also consult \cite{MinMooTsi}).  Because
we have compactified the spatial dimension on $S^1$, we are dealing
with a closed string with left- and right-moving states. The 4
supersymmetries of the conformal field theory reside in one of these
two sectors, say the right-handed one.  Conformal theories in $1+1$
dimensions are characterized by a central charge, and in this case
there is a central charge for the right- and for the left-moving sector
separately. These central charges are expressible in terms of the
wrapping numbers $p^A$ and depend on the intersection numbers and the
second Chern class, according to
\begin{eqnarray}
  \label{eq:c-LR}
  c_L &=& C_{ABC}\, p^Ap^Bp^C + c_{2A}\,p^A \,,\nonumber \\
  c_R &=& C_{ABC}\, p^Ap^Bp^C + \tfrac12  c_{2A}\,p^A \,.
\end{eqnarray}
We should stress that the above result is far from obvious and holds
only under the condition that the $p^A$ are large. 
In that case every generic deformation of ${\cal P}$ will be smooth.
Under these circumstances it is possible to relate the topological
properties of the 4-cycle to the topological data of the Calabi-Yau
space. 

We can now choose a state of given momentum $q_0$ which is supersymmetric
in the right-moving sector. From rather general arguments it follows that
such states exist. The corresponding states in the left-moving sector
have no bearing on the supersymmetry and these states have a certain
degeneracy depending on the value of $q_0$. In this way we have a
tower of BPS states invariant under 4 supersymmetries, built on
supersymmetric states in the right-moving sector and comprising
corresponding degenerate states in the left-moving sector.  We can
then use Cardy's formula, which states that the degeneracy of states
for fixed but large momentum (large as compared to $c_L$) equals
$\exp[2\pi\sqrt{\vert q_0\vert \, c_L/6}]$.  This leads to the
following expression for the entropy,
\begin{equation}
  \label{eq:S-CY-micro}
  {\cal S}_{\rm micro}(p,q) = 2\pi\sqrt{ \tfrac16 
  \vert\hat q_0\vert (C_{ABC} \,p^Ap^Bp^C + c_{2A}\,p^A)}\,,
\end{equation}
where $q_0$ has been shifted according to
\begin{equation}
  \label{eq:hat-q}
  \hat q_0 = q_0 + \tfrac1{2} C^{AB} q_Aq_B\,.
\end{equation}
Here $C^{AB}$ is the inverse of $C_{AB}= C_{ABC}p^C$.  This
modification is related to the fact that the electric charges
associated with the gauge fields $A_\mu{}^A$ will interact with the
M-theory two-brane \cite{Maldacena:1997de}. The existence of this
interaction can be inferred from the fact that the two-brane interacts
with the rank-three tensor field in eleven dimensions, from which the
vector gauge fields $A_\mu{}^A$ originate.

We stress that the above results apply in the case of large charges.
The first term proportional to the triple intersection number is
obviously the leading contribution whereas the terms proportional to the
second Chern class are subleading. The importance of the subleading
terms will become more clear in later sections. Having obtained a
microscopic representation of a BPS black hole, it now remains to make
contact with it by deriving the corresponding black hole solution
directly in the $N=2$ supergravity theory. This is discussed in the
next section.

\section{Entropy formula for $N=2$ supergravity}
\label{sec:entropy-formula}
The charged black hole solutions in $N=2$ supergravity are invariant
under 4 of the 8 supersymmetries. They are solitonic, and interpolate
between fully supersymmetric configurations at the horizon and at
spatial infinity. At spatial infinity, where the effect of the charges
can be ignored, one has flat Minkowski space-time. The scalar moduli
fields tend to certain (arbitrary) values on which the black hole mass
will depend.  At the horizon the situation is rather
different, because here the charges are felt and one is not longer
dealing with flat space-time, but with a so-called Bertotti-Robinson
space, ${\rm AdS}_2\times S^2$.  In that situation the requirement of
full $N=2$ supersymmetry is highly restrictive and for spherical
geometries one can prove that the values of the moduli fields at the
horizon are in fact fixed in terms of the charges. The corresponding
equations are known as the attractor equations
\cite{Ferrara:1995ih,Strominger:1996kf, Ferrara:1996dd} and they apply
quite generally. For effective actions with interactions quadratic in
the curvature the validity of these attractor equations was
established in \cite{LopesCardoso:2000qm}.

The attractor equations play a crucial role as they ensure that the
entropy, a quantity that is associated with the horizon, will depend
on the black hole charges and not on other quantities, in line with
the microscopic results presented in the previous section. Hence we are
interested in studying charged black hole solutions which are
BPS, meaning that they are invariant under 4 supersymmetries. The
matter supermultiplets contain the gauge fields $A_\mu{}^A$ coupling
to electric and magnetic charges $q_A$ and $p^A$, respectively. In
addition there is one extra graviphoton field $A_\mu{}^0$ which may
couple to charges $q_0$ and $p^0$. When comparing to the solutions of
the previous section, we obviously will set $p^0=0$, but from the
supergravity point of view there is no need for such a restriction.

However, there is an infinite variety of $N=2$ supergravity actions
coupling to vector multiplets. Fortunately these actions can be
conveniently encoded into holomorphic functions that are homeogeneous
of second degree \cite{DWVHVPL}. In the case at hand the simplest
action is, for instance, based on the function,
\begin{equation}
  \label{eq:CY-sg}
  F(Y)=-\frac1{6}  \frac{C_{ABC}\,Y^AY^BY^C}{Y^0} \,,
\end{equation}
where the holomophic variables $Y^I$ ($I=0,A$) are associated with the
vector multiplets; they can be identified projectively with the scalar
moduli fields that are related to a subset of the moduli of the
Calabi-Yau space.  The black hole solution will thus encode the changes in the 
Calabi-Yau manifold when moving from the black hole horizon
towards spatial infinity, precisely as discussed in a more general
context in section~3. Note the presence of the triple intersection
form $C_{ABC}$ which will appear in the interaction vertices of the
corresponding Lagrangian. 

The attractor equations also involve the function $F(Y)$. In terms of
the quantities $Y^I$ and the first derivatives of $F(Y)$, they take
the form,
\begin{equation}
  \label{eq:attractor}
  Y^I-\bar Y^I = \mathrm{i} p^I\,,\qquad 
  F_I(Y) - \bar F_I(\bar Y) = \mathrm{i} q_I\,,
\end{equation}
where $F_I(Y)= \partial F(Y)/\partial Y^I$. In principle these
equations yield the horizon values of the $Y^I$ in terms of the
charges. Depending on the values of the charges and on the complexity
of the function $F(Y)$, it may not be possible to write down
solutions in closed form.

The action corresponding to (\ref{eq:CY-sg}) gives rise to a black
hole solution with charges $p^A$, $q_A$ and $q_0$ (we take $p^0=0$).
Its area can be calculated and is equal to
\begin{equation}
  \label{eq:S-CY-macro}
  A(p,q) = 8 \pi\sqrt{ \tfrac16 
  \vert\hat q_0\vert \,C_{ABC} \,p^Ap^Bp^C }\,.
\end{equation}
Upon invoking the area law this result leads precisely to the first
part of the microscopic entropy (\ref{eq:S-CY-micro}). We have thus
reproduced the leading contributions to the entropy from supergravity.

How can one reproduce the subleading terms in view of the fact that
these terms scale differently whereas the function $F(Y)$ and the
attractor equations all seem to scale uniformly? To explain how this
is resolved we must first spend a few words on the reason why the
function $F(Y)$ was homogeneous in the first place. The covariant
fields corresponding to a vector supermultiplet comprise a so-called
restricted chiral multiplet, which can be assigned a unique (complex)
scaling weight. The $Y^I$ are proportional to the lowest component of
these multiplets, and can be assigned the same scaling weight. Any
(holomorphic) function of these restricted multiplets will define a
chiral superfield, whose chiral superspace integral will lead to a
supersymmetric action.  However, in order to be able to couple to
supergravity, this function must be homogeneous of second degree
\cite{DWVHVPL}.  To deviate from this homogeneity pattern in the
determination of the entropy, one needs to introduce a new type of
chiral superfield whose value at the horizon will be fixed in a way
that breaks the uniformity of the scaling.  There exists such a
multiplet.  Namely, from the fields of (conformal) supergravity
itself, one can again extract a restricted chiral multiplet, which in
this case comprises the covariant quantities associated with the
supergravity fields.  This time the restricted chiral multiplet is not
a scalar, but an auxiliary anti-selfdual tensor, and just as before it
can be assigned a unique scaling weight. Its lowest component is an
auxiliary field that is often called the graviphoton field strength,
which is strictly speaking a misnomer because it never satisfies a
Bianchi identity. It appears in the transformation rule of the
gravitino fields and in simple Lagrangians its field equations
express it in terms of moduli-dependent linear combinations of other
vector field strengths.  The square of this restricted tensor defines
a complex scalar field which constitutes the lowest component of a
chiral supermultiplet and which is proportional to a field we will
denote by $\Upsilon$, such that the (complex) scaling weight that can be
assigned to $\Upsilon$ is twice that of the $Y^I$. More general
supergravity Lagrangians can then be described by holomorphic functions
$F(Y,\Upsilon)$ that are homogeneous of degree 2, {\it i.e.},
\begin{equation}
  \label{eq:homogeneous-F}
  F(\lambda Y^I,\lambda^2 \Upsilon) = \lambda^2\, F( Y^I,\Upsilon)\,. 
\end{equation}
A nontrivial dependence on $\Upsilon$ in the function $F(Y,\Upsilon)$
has important consequences, because the supermultiplet of which
$\Upsilon$ is the first component contains other components with
terms quadratic in the Riemann tensor.  Hence actions based on a
function (\ref{eq:homogeneous-F}) with a nontrivial dependence on
$\Upsilon$ will contain terms proportional to the square of the
Riemann tensor, multiplied by the first derivative of $F$ with respect
to $\Upsilon$.  The attractor equations (\ref{eq:attractor}) remain
valid with $F(Y)$ replaced by $F(Y,\Upsilon)$. However, the field
$\Upsilon$ has its own independent attractor value; at the horizon it 
must be equal to $\Upsilon=-64$, independent of the charges.  This
phenomenon explains why the area and the entropy are not
necessarily a homogeneous function of the charges.

To fully reproduce the entropy formula (\ref{eq:S-CY-micro}) including
the subleading terms proportional to the second Chern class, one may
attempt a Lagrangian based on the function
\begin{equation}
  \label{eq:CY+cc-sg}
  F(Y)=-\frac1{6} \, \frac{C_{ABC}\,Y^AY^BY^C}{Y^0} - \frac{c_{2A}} 
   {24\cdot 64}\, \frac{ Y^A}{Y^0} \,\Upsilon \,,
\end{equation}
which is indeed holomorphic and homogeneous. On the basis of this
modification one can again calculate the horizon area in the hope of
recovering the entropy (\ref{eq:S-CY-macro}) upon use of the area law.
However, the result is negative and it seems obvious that no solution
can be found in this way \cite{BCDWLMS}. 

At this point the only way out is to no longer rely on the area law in
extracting a value for the entropy. Indeed the area law is not
expected to hold for actions that supersede the Einstein-Hilbert one.
Wald has proposed an alternative definition of black hole
entropy which can be used for any Lagrangian that is invariant under
general coordinate transformations, and which is based on the
existence of a conserved surface charge \cite{WaldIyer}. The latter is
related to the conventional Noether current associated with general
coordinate transformations, which, for a gauge symmetry, can be
written as a pure improvement term: the divergence of an antisymmetric
tensor, called the Noether potential. It turns out that with the help
of the Noether potential one can define a surface charge integrated
over the boundary of a Cauchy surface, which for the black hole
extends from spatial infinity down to the horizon. Changes in the
continuous variety of black hole solutions should leave this charge 
unchanged.  Under certain conditions one can show that the change of
the surface integral at spatial infinity corresponds to the mass and
angular momentum variations in the first law. Therefore one identifies
the surface integral at the horizon with the entropy, so that the
validity of the first law will remain ensured.

We should stress that there are various subtle points here, some of which 
have been discussed in \cite{CarDeWMoh}.  The prescription based
on the surface charge can be applied to standard Einstein gravity, in
which case one just recovers the area law. But for theories with
higher-derivatives, there are nontrivial correction terms, which
follow from a calculation of the Noether potential. We should caution
the reader that the relevant correction term in the case at hand does
in fact not reside in the terms quadratic in the Riemann tensor, but
in some other terms related to them by supersymmetry.

From an evaluation of the Noether potential, taking into account all
the constraints imposed by the supersymmetry at the horizon, it 
follows that the entropy can be written in a universal form
\cite{LopesCardoso:1998wt},
\begin{equation}
  \label{eq:W-entropy}
  {\cal S}_{\rm macro} (p,q) = \pi \Big[\vert Z\vert^2 - 256 \, 
  {\rm Im} F_\Upsilon \Big ]_{\Upsilon=-64} \,. 
\end{equation}
Here the first term denotes the Bekenstein-Hawking entropy, because
$\vert Z\vert^2= p^IF_I(Y,\Upsilon)- q_IY^I$ is just the area in
Planck units divided by $4\pi$. This term is clearly affected by the
presence of the higher-order derivative interactions. On top of that
there is a second term proportional to the derivative of
$F(Y,\Upsilon)$ with respect to $\Upsilon$. This term
thus represents the deviation of the area law. The above formula applies
to any $N=2$ supergravity solution.

Because of the fact that all quantities of interest are directly
related to the holomorphic and homogeneous functions
(\ref{eq:homogeneous-F}), the determination of the area and entropy is
merely an algebraic exercise, which no longer requires to construct
the full solution. Given the function $F(Y,\Upsilon)$ one first
attempts to solve the attractor equations (\ref{eq:attractor}), but
now with $F_I(Y)$ replaced by $F_I(Y,\Upsilon)$. Subsequently one determines
the area and the entropy in terms of the charges. For the function
(\ref{eq:CY+cc-sg}) this was shown to lead precisely to the
microscopic entropy formula (\ref{eq:S-CY-micro}). To exhibit the
deviation from the area law, we also give the area (which is obviously
not known from microscopic considerations),
\begin{equation}
  \label{eq:area/entropy}
 \tfrac1{4} A(p,q) = \frac{C_{ABC}\, p^Ap^Bp^C + \tfrac12 c_{2A}\,p^A}
  {C_{ABC}\, p^Ap^Bp^C + c_{2A}\,p^A} \;
 {\cal S}_{\rm macro} (p,q) \,. 
\end{equation}
Interestingly the proportionality factor is just the ratio
$c_R/c_L$ of the two central charges defined in (\ref{eq:c-LR}). By
combining new ingredients from supergravity and general relativity it
is thus possible to fully account for the black hole entropy that is
obtained by counting microstates.

Before moving to the next section we wish to add some observations. In
addition to being able to evaluate the properties of the black hole
solution at the horizon one should also like to understand the full
structure of the BPS black holes away from the horizon, in the
presence of the interactions quadratic in the Riemann curvature. This
was the subject of \cite{LopesCardoso:2000qm}, where a rather general
class of such solutions was studied, including multi-centered ones.
We refer to that work for further details. It is also worth
pointing out that we have always been basing ourselves on the 
effective Wilsonian action. A priori, one does
not expect that the final macroscopic description of black hole
mechanics can be obtained exclusively within a Wilsonian framework. 

\section{Heterotic black holes}
\label{sec:heter-black-holes}
In \cite{CarDeWMoh,LopesCardoso:1999ur} the modified entropy formula
(\ref{eq:W-entropy}) was applied to heterotic black holes. Although
the formula is derived for $N=2$ supergravity, the result can readily
be generalized to the case of heterotic $N=4$ supersymmetric
compactifications. This involves an extension of the target-space
duality group to ${\rm SO}(6,22)$ with a corresponding extension to 28
electric and 28 magnetic charges that take their values in a
$\Gamma^{6,22}$ lattice. The $N=4$ supersymmetric heterotic models
have dual realizations as type-II string compactifications on
$K3\times T^2$. In contrast to $N=2$ Calabi-Yau compactifications, the
holomorphic function which encodes the effective Wilsonian action is
severely restricted in the $N=4$ case.  Therefore it is often possible
to obtain exact predictions in this context.

The relevant function for the heterotic case takes the following
form in lowest order,
\begin{equation}
  \label{eq:F-het}
  F(Y) = - \frac{Y^1\,Y^a\eta_{ab}Y^b}{Y^0} \,,
\end{equation}
where $a,b= 2,\ldots,n$. This function will be modified in due course
by a function of $\Upsilon$ and of the dilaton field $S=-\mathrm{i} Y^1/Y^0$.
Let us fist consider (\ref{eq:F-het}) in the absence of these modifications. 
Then the $2n$ scalar moduli in the effective action are described by a
nonlinear sigma model with the following target space,
\begin{equation}
  \label{eq:heterotic-duality}
  {\cal M} = \frac{{\rm SU}(1,1)}{{\rm U}(1)} \times \frac{{\rm SO}(2,n-1)}
   {{\rm SO}(2)\times {\rm SO}(n-1)} \,. 
\end{equation}
The electric and magnetic charges transform under the action of the
${\rm SU}(1,1)\times{\rm SO}(2,n-1)$ isometry group. The first factor,
${\rm SU}(1,1)$, is associated with $S$-duality. This is a strong-weak
coupling duality which interchanges electric and magnetic charges. The
second factor, ${\rm SO}(2,n-1)$, is associated with $T$-duality (also
called target-space duality). There is a technical complication in the
description based on (\ref{eq:F-het}) because the charges that follow
from the $Y^I$ through the attractor equations are not in a convenient
basis for $S$- and $T$-duality. A proper basis is found upon
interchanging the electric and magnetic charges $q_1$ and $p^1$ by
an electric/magnetic duality. Fortunately there is no need to discuss
this in any detail, as the entropy and area of the corresponding black
holes depend only on $T$-duality invariants of the charges. Note that
in the extension to $N=4$ the second factor of the target space
(\ref{eq:heterotic-duality}) changes into ${\rm SO}(6,22) /[{\rm 
  SO}(6)\times {\rm SO}(22)]$; this space is parametrized by the 132
scalar fields belonging to 22 $N=4$ vector supermultiplets.

To be specific let us first quote the result for the entropy and
horizon area for the solution based on (\ref{eq:F-het}) as a function
of the charges,
\begin{equation}
  \label{eq:class-het-entropy}
  {\cal S}_{\rm macro}(p,q)= \tfrac14 A(p,q)  
  = \pi \sqrt{q^2\, p^2 -(p\cdot q)^2} \,.
\end{equation}
Here we used $T$-duality invariant combinations of the charges, defined by 
\begin{eqnarray}
  \label{eq:charge-invariants}
  q^2&=& 2\,q_0p^1 - \tfrac12 q_a\eta^{ab} q_b\,, \nonumber \\
  p^2 &=& -2 p^0q_1 - 2\,p^a\eta_{ab}p^b\,, \nonumber \\
  p\!\cdot \!q &=& q_0p^0-q_1p^1 + q_ap^a \,,
\end{eqnarray}
where the $p^I$ and $q_I$ on the right-hand side are the charges that
appear in the attractor equations (\ref{eq:attractor}) based on the
function (\ref{eq:F-het}). While the combinations
(\ref{eq:charge-invariants}) are invariant under $T$-duality, they
transform as an ${\rm SO}(2,1)$ vector under $S$-duality, such that
the entropy formula (\ref{eq:class-het-entropy}) is invariant under
both $T$- and $S$-duality. 
The $S$-duality transformations, which constitute the group
$\mathrm{SL}(2,\mathbb{Z})$, of the charges and of the dilaton field
are related through the attractor equations. They are given by
\begin{eqnarray}
  \label{eq:S-duality}
  S &\longrightarrow& \frac{a\,S -\mathrm{i} b}{\mathrm{i}c\,S +d}\,, 
\nonumber\\ 
   q^2&\longrightarrow& a^2\,q^2 + b^2\, p^2 + 2\,ab\, p\!\cdot\! q \,,
\nonumber\\
   p^2&\longrightarrow& c^2\,q^2 + d^2\, p^2 + 2\,cd\, p\!\cdot\! q \,,
\nonumber\\
   p\!\cdot\! q&\longrightarrow& ac\,q^2 + bd\, p^2 + (ad+bc)\,
   p\!\cdot\! q \,, 
\end{eqnarray}
with integer-valued parameters $a,b,c,d$ satisfying $ad-bc=1$. In the
$N=4$ theory, the charge lattice is $S$-duality invariant, meaning
that the above transformations always lead to a point on the
lattice that is physically realized. Note that the supergravity calculations
yield no intrinsic definition of the normalization of the charge
lattice and consequently the dilaton normalization is a priori not
known. Hence the precise characterization of the arithmetic subgroup
of $\mathrm{SL}(2)$ that defines the $S$-duality group is not
obvious, but the crucial point is that the normalization of the
dilaton is related to the normalization of the lattice of charges.
Later on in this section we will relate the supergravity results to
microscopic data which will confirm the above identifications. Observe
that the invariants $p^2$ and $q^2$ are not positive definite. In fact
in the limit of large charges they will both become negative.

When adding a function to (\ref{eq:F-het}) proportional to $\Upsilon$
and depending otherwise on the dilaton field $S$, it turns out that
the target-space duality remains unaffected. However, $S$-duality is
affected in general so the question is whether there exists a
specific modification that leaves $S$-duality intact. This turns out
to be the case, but one is forced to accept a certain amount of
non-holomorphicity in the description \cite{LopesCardoso:1999ur}.  To
derive this is rather nontrivial and we simply quote the result in a
form that was indicated in \cite{CDWKM},
\begin{equation}
  \label{eq:F-het-mod}
  F(Y,\bar Y,\Upsilon,\bar\Upsilon) = - \frac{Y^1\,Y^a\eta_{ab}Y^b}{Y^0} +  
  \frac{\mathrm{i}}{64\pi} 
   \,\Upsilon \log\eta^{12}(S)  +\frac{\mathrm{i}}{128\pi}\, 
  (\Upsilon+\bar \Upsilon)   \log (S+\bar S)^6 \,,
\end{equation}
where the non-holomorphic corrections reside in the last term.
For the convenience of the reader we define the Dedekind eta-function, 
\begin{equation}
  \label{eq:Dedekind}
  \eta(q_\tau) = q_\tau^{\,1/24} \, \prod_{n=1}^{\infty} (1-q_\tau^{\,n}) \,,
\end{equation}
where $q_\tau = \exp (2\pi\mathrm{i}\tau)$. It satisfies the
asymptotic formula, $\ln \eta(q_\tau) = \tfrac1{24}\ln q_\tau - q_\tau
- \tfrac32 q_\tau^{\,2} - \tfrac43 q_\tau^{\,3} - \tfrac74
q_\tau^{\,5} + {\cal O} (q_\tau^{\,6})$. By $\eta(S)$ we mean
$\eta(q_\tau)$ with $\tau= \mathrm{i} \,S$, so that $\log
\eta(S)\approx -\frac1{12} \pi\,S - {\mathrm e}^{-2\pi S}+ {\cal
  O}({\mathrm e}^{-4\pi S})$. We also recall that $\eta^{24}(S)$ is a
modular form of degree 12, so that $\eta^{24}(S^\prime) = (\mathrm{i}
\, c\,S+d)^{12}\,\eta^{24}(S)$, where $S^\prime$ is the transformed
dilaton field as defined in (\ref{eq:S-duality}).

The presence of non-holomorphic terms in (\ref{eq:F-het-mod}) is not
entirely unexpected: the Wilsonian couplings are holomorphic but may
not fully reflect the symmetries of the underlying theory, while the
physical couplings must reflect the symmetry and may thus have
different analyticity properties. The non-holomorphic terms are
determined uniquely by requiring $S$-duality and consistency with
string perturbation theory, and are in accord with the $N=4$ results
of \cite{Harvey:1996ir}. The normalization of the new terms is,
however, not fixed by $S$-duality. It has been fixed by using
string-string duality, or, alternatively, by requiring agreement with
the known asymptotic degeneracy of electrically charged black holes,
as we shall see later on.

Including the non-holomorphic corrections, the result of
\cite{LopesCardoso:1999ur} can be summarized as follows. The
non-trivial attractor equations are the ones that determine the
horizon value of the complex dilaton field $S$ in terms of the black
hole charges. They read as follows (we have now set $\Upsilon$ to its
horizon value),
\begin{eqnarray}
\label{eq:nonholostab}
\vert S\vert^2 \, p^2 &=& q^2 -  \frac{ 2}{\pi} ( S + {\bar S}) \, 
\Big(S \frac{\partial}{\partial S}+ \bar S \frac{\partial}{\partial \bar
  S}\Big) 
 \log\left[ (S+\bar S)^6\vert\eta(S)\vert^{24}\right] \;, \nonumber\\ 
(S - {\bar S}) \,p^2 &=& {}-2 \,\mathrm{i} \, p\!\cdot \!q + \frac{2}{\pi} 
 (S + {\bar S}) \, \Big(\frac{\partial}{\partial S} 
 - \frac{\partial}{\partial\bar S}\Big) 
\log\left[ (S+\bar S)^6\vert \eta(S)\vert^{24}\right] \;.
\end{eqnarray}
The expression for the macroscopic entropy reads, 
\begin{eqnarray}
  \mathcal{S}_\text{macro} = 
- \pi \left[ \frac{q^2 - \mathrm{i} p\!\cdot\! q \, 
(S - {\bar S}) + p^2 \,|S|^2} 
{S + {\bar S}} \right] -2 \, \log\left[ (S + {\bar S})^6
  |\eta(S)|^{24}\right] \;, 
\label{eq:nonholoentropy}
\end{eqnarray}  
with the dilaton field subject to (\ref{eq:nonholostab}). The first
term in this equation corresponds to one-fourth of the horizon area,
which, via (\ref{eq:nonholostab}), is affected by the various
corrections.  The second term represents an extra modification, which
explicitly contains the non-holomorphic correction. Both terms are
invariant under target-space duality and $S$-duality. As explained
above, $S$-duality was achieved at the price of including non-holomorphic
terms, here residing in the $\log(S+\bar S)$ terms.

In string perturbation theory the real part of $S$ becomes large and
positive, and one can neglect the exponential terms of the Dedekind
eta-function. In that approximation the imaginary part of $S$ equals
${\rm Im}\,S= - p\!\cdot\!q/p^2$ and the 
real part is determined by a quadratic equation,
\begin{equation}
  \label{eq:re-S}
  \tfrac1{4} p^2 \left(p^2 -8\,\right)
\, (S+\bar S)^2 +  \frac{12}{\pi}\, p^2  
  \,(S+\bar S) = {q^2\,p^2 -(p\!\cdot\!q)^2} \,.
\end{equation}
Obviously, these perturbative results are affected by the presence of
the non-holomorphic corrections.  Using (\ref{eq:re-S}), we find the
following expression for the corresponding entropy,
\begin{equation}
\label{eq:entcor}
  \mathcal{S}_{\text{macro}} = - 
{2\pi}  \, \frac {[q^2 \, p^2 - (p\!\cdot\!q)^2]}{p^2\,(S+\bar S)} 
 -12 \left[\log (S+\bar S) - 1\right] \,.
\end{equation}
For large charges and finite value of $S$, the above result for ${\cal
  S}_{\text{macro}}$ tends to (\ref{eq:class-het-entropy}).

In the $N=4$ setting one can distinguish two types of BPS-states.
Purely electric or magnetic configurations constitute 1/2-BPS states,
whereas dyonic ones are 1/4-BPS states. For $N=2$ the distinction
between the two types of states disappears and one has only 1/2-BPS
states. In the context of $N=4$, the generic BPS states are the dyonic
ones, characterized by a nonzero value for $q^2\,p^2
-(p\!\cdot\!q)^2$.  Actually, the $S$-duality invariant
characterization of the 1/2-BPS states, is precisely expressed by the
condition $q^2\,p^2-(p\!\cdot\!q)^2=0$.  In the remainder of this
section we restrict our attention to the dyonic states. The
macroscopic results given above can be confronted with an explicit
formula for the microscopic degeneracy of BPS dyons in
four-dimensional $N=4$ string theory proposed in
\cite{Dijkgraaf:1996it}. This proposal generalizes the expression for
the degeneracies of electric heterotic string states (to be presented
in the next section), to an expression that depends on both electric
and magnetic charges such that it is formally covariant with respect
to $S$-duality. In \cite{Dijkgraaf:1996it} it was already shown that
the dyonic degeneracy was consistent with the area law, {\it i.e.}
with (\ref{eq:class-het-entropy}) in the limit of large charges.

Let us be a little more specific about the actual degeneracy formula. It is
expressed in terms of an integral over an appropriate 3-cycle that
involves an automorphic form $\Phi_{10}(\Omega)$,
\begin{equation}
  \label{eq:dvvdeg}
  d(q, p) = \oint {\mathrm d} \Omega \,
 \frac{{\mathrm e}^{\mathrm{i}\pi (Q^T \Omega \,Q)}}{\Phi_{10}(\Omega)} \;.
\end{equation}
Here $\Omega$ denotes the period matrix for a genus-2 Riemann surface,
which parametrizes the $\mathrm{Sp}(2)/\mathrm{U}(2)$ cosets; it 
can be written as a complex, symmetric, two-by-two matrix,
\begin{equation}
  \label{eq:oq}
  \Omega = 
  \begin{pmatrix} 
    \rho & \upsilon \\ \upsilon &\sigma 
  \end{pmatrix}\,.
\end{equation}
In the exponent of the numerator of (\ref{eq:dvvdeg}) the direct
product of the period matrix with the invariant metric of the charge
lattice is contracted with the charge vector comprising the 28
magnetic and 28 electric charges, so that $Q^T \Omega \,Q = \rho\, p^2
+ \sigma\,q^2 +2\, \upsilon\,p\!\cdot\!q$, where $p^2$, $q^2$ and
$p\!\cdot\! q$ were defined previously in
(\ref{eq:charge-invariants}). A representation of $\Phi_{10}$ in the
form of a Fourier series with three complex arguments is, for
instance, given in \cite{Gritsenko:1995xxx},
\begin{eqnarray}
    \label{eq:product-form}
   \Phi_{10}(\Omega) =  q_\rho q_\sigma q_\upsilon 
   \prod_{\{k,l,m\}} (1- q_\rho^{\,k}\, q_\sigma^{\,l}\, 
    q_\upsilon^{\,m})^{c(kl,m)}\,,    
\end{eqnarray}
where $q_\rho= \exp(2\pi\mathrm{i}\rho)$, $ q_\sigma =
\exp(2\pi\mathrm{i}\sigma)$ and $q_\upsilon =
\exp(2\pi\mathrm{i}\upsilon)$. Some more details can be found in
\cite{CDWKM}. The inverse of $\Phi_{10}$ has poles and the
formula~(\ref{eq:dvvdeg}) picks up a corresponding residue whenever
the 3-cycle encloses such a pole. However, the poles are located in
the interior of the Siegel half-space and not just at its boundary and
therefore the choice of the 3-cycles is subtle.

In \cite{CDWKM} the degeneracy formula was studied by means of a
saddle-point approximation in the limit of large charges, but now
retaining also the terms that are subleading. Remarkably enough the
result is in precise agreement with the macroscopic results, {\it
  i.e.} with (\ref{eq:nonholostab}) and (\ref{eq:nonholoentropy}),
including the non-holomorphic terms. The equations
(\ref{eq:nonholostab}) turn out to correspond to the equations that
determine the location of the saddle-point, while
(\ref{eq:nonholoentropy}) represents the value of the integrand in
(\ref{eq:dvvdeg}) taken at the saddle-point, including the contribution
from integrating out the fluctuations about the saddle-point. This
shows that the macroscopic entropy, defined by
(\ref{eq:nonholoentropy}) as a function of the charges and the dilaton
field, is in fact stationary under variations of the latter. In fact,
it is possible to understand this stationarity principle on the basis
of the variational principle proposed in \cite{BCDWKLM}, upon a proper 
extension with $\Upsilon$-dependent terms and non-holomorphic
corrections.

\section{The area law and elementary string states}
\label{sec:area}
The area law is clearly violated in the presence of the subleading
corrections, as is shown in the $N=2$ entropy
formula~(\ref{eq:W-entropy}). Of course, it depends on the theory in
question and on the values for the charges, how sizable this violation
is.  A particularly interesting case emerges for black holes for which
the leading contribution to the entropy and area vanish. In that case,
the subleading terms become dominant and (\ref{eq:area/entropy}) shows
that the area law is replaced by ${\cal S}_{\text{macro}}= \tfrac12
A(p,q)$, whereas the typical dependence on the charges proportional to
the square root of a quartic polynomial is changed into the square
root of a quadratic polynomial. It is easy to see how this can be
accomplished for the heterotic black holes, namely, by suppressing all
the charges in (\ref{eq:charge-invariants}) with the exception of
$q_0$ and $p^1$, leading to $q^2$ as the only nonvanishing $T$-duality
invariant charge combination.  This is a remarkable result. In fact these
states are precisely generated by perturbative heterotic string states
arising from a compactification of six dimensions. In the
supersymmetric right-moving sector they carry only momentum and
winding and contain no oscillations, whereas in the left-moving sector
oscillations are allowed that satisfy the string matching condition.
The oscillator number is then linearly related to $q^2$. These string
states are 1/2-BPS states and correspond to electrically charged
states (possibly upon a suitable electric/magnetic duality
redefinition).

Precisely these perturbative states already received quite some attention
in the past (for an early reference, see \cite{DaHa}). Because the
higher-mass string BPS states are expected to be within their
Schwarzschild radius, it was conjectured that they should have an
interpretation as black holes.\footnote{ 
  The idea that elementary particles, or string states, are behaving
  like black holes, has been around for quite some time. However, it
  is outside the scope of these lectures to discuss this in more
  detail.   } 
Their calculable level density, proportional to the exponent of
$4\pi \sqrt{\vert q^2\vert/2 }$, implies a nonzero microscopic entropy
for these black holes \cite{Russo:1994}. On the other hand the
corresponding black hole solutions were constructed in
\cite{Sen:1994,Sen:1995in} and it was found that their horizon area
vanishes, which, on the basis of the area law, would imply a vanishing
macroscopic entropy.

Of course, higer-order string corrections are expected to modify the
situation at the horizon. One of the ways to incorporate their effect
is to make use of the concept of a `stretched' horizon, a surface
close to the event horizon whose location is carefully adapted 
in order that the calculations remain internally consistent. In this
way it is possible to reconcile the non-zero level density with the
vanishing of the classical horizon area \cite{Sen:1995in,Peet:1995},
although the precise proportionality factor in front of $\sqrt{\vert
  q^2\vert/2 }$ cannot be determined.

On the other hand, these corrections will undoubtedly be related to
interactions of higher order in the curvature tensor whose effect can be
studied in the context of the modified entropy formula (\ref{eq:W-entropy})
together with the attractor equations (after all, their derivation did
not pose any restriction on the values of the leading contributions
to area and entropy). In fact, some of the results can be read off
easily from the formulae (\ref{eq:nonholostab}) and
(\ref{eq:nonholoentropy}).  They show that the dilaton field becomes
large and real (in contrast with the dyonic case, where the dilaton
could remain finite and complex). Direct substitution yields,
\begin{eqnarray}
\label{eq:entcor-electric}
  S+\bar S &\approx &\sqrt{\vert q^2\vert/2}\,,\nonumber \\ 
  \mathcal{S}_{\text{macro}} &\approx& 
4\,\pi\, \sqrt{\vert q^2\vert/2} -6\, \log{\vert q^2\vert}  \,, 
\end{eqnarray}
where the logarithmic term is due to the non-holomorphic contribution.
Because the dilaton is large in this case, all the exponentials in the
Dedekind eta-function are suppressed and we are at weak string
coupling. Consequently the formalism discussed in
section~\ref{sec:entropy-formula} yields the expected results. In
fact, without the non-holomorphic corrections the result for the
entropy can be obtained directly from (\ref{eq:S-CY-micro}), upon
taking $\hat q_0=q_0$, $C_{ABC}= 0$ and $c_{2\,A}\,p^A = 24\, p^1$.

Not much attention was paid to this particular application of
(\ref{eq:W-entropy}) until recently, when attention focused again
on the electric black holes \cite{Dabholkar:2004yr}, this time
primarily motivated by a reformulation of the black hole entropy
(\ref{eq:W-entropy}) in terms of a mixed partition function
\cite{Ooguri:2004zv}. We turn to the latter topic in the next section.
The observation that (\ref{eq:W-entropy}) can nicely account for the
discrepancies encountered in the classical description of the 1/2-BPS
black holes was first made in
\cite{Dabholkar:2004yr,Dabholkar:2004dq}. Note also that, since the
electric states correspond to perturbative heterotic string states,
their degeneracy is known from string theory and given by
\begin{equation}
  \label{eq:het-string}
  d(q) = \oint {\mathrm d}\sigma\, \frac{{\mathrm
  e}^{\mathrm{i}\pi\sigma q^2}}{\eta^{24}(\sigma)} \approx 
  \exp\left(4\pi\,\sqrt{|q^2|/2} - \tfrac{27}{4}
      \log \vert q^2\vert\right)  \,,
\end{equation}
where the integration contour encircles the point $\exp(2\pi
\mathrm{i} \sigma)=0$. The large-$\vert q^2\vert$ approximation is
based on a standard saddle-point approximation.  Obviously the
leading term of (\ref{eq:het-string}) is in agreement with
(\ref{eq:entcor-electric}). However, the logarithmic corrections carry
different coefficients.
 
At this point we should recall that the dyonic degeneracy formula
(\ref{eq:dvvdeg}) was proposed at the time \cite{Dijkgraaf:1996it} as
an $S$-duality invariant extension of the electric degeneracy formula
(\ref{eq:het-string}). While for the dyonic states the macroscopic
results agree fully with the results obtained from a saddle-point
approximation of (\ref{eq:dvvdeg}), we conclude that the situation
regarding the electric states is apparently more subtle.  We refrain
from discussing this in more detail.

Recently, there have been quite a number of papers about the electric
black holes, discussing the effect of the higher-derivative
corrections in the effective action on the horizon behaviour and on
more global aspects of the black hole solutions
\cite{Sen:2004dp,Hubeny:2004ji,Bak:2005x,Sen:2005x}; two of them also
discuss the effect of the non-holomorphic corrections
\cite{Sen:2004dp,Sen:2005x}. Other papers are pursueing the
consequences of the conjecture of \cite{Ooguri:2004zv}, where a mixed
black hole partition function is proposed proportional to the square
of the topological string partition function
\cite{Dabholkar:2004yr,Dabholkar:2005x}. We turn to a discussion of
this partition function and some of its consequences in the next
section.

\section{A black hole partition function}
\label{sec:black-hole-partition}
The attractor equations
\cite{Ferrara:1995ih,Strominger:1996kf,Ferrara:1996dd} were originally
interpreted as the conditions that extremized the central charge ({\it
  i.e.}, the so-called BPS charge $Z$, as at the horizon there is full
supersymmetry so that the supersymmetry algebra will not exhibit a
central charge). Subsequently a variational principle was written down
in \cite{BCDWKLM} for some `potential' ${\cal V}$, which was indeed
stationary whenever the attractor equations were satisfied; at the
stationary point ${\cal V}$ was precisely equal to the macroscopic
entropy (up to some normalization). This variational principle did not
receive much attention, but it recently it re-emerged in the work of
\cite{Ooguri:2005x}. Meanwhile, in a separate development, another
variational principle had been introduced \cite{Ooguri:2004zv}, and
furthermore, as we already noted at the end of
section~\ref{sec:heter-black-holes}, the attractor equations and the
entropy for heterotic black holes, expressed in terms of the charges
and the dilaton field, seem to be based on an underlying variational
principle. It is unlikely that all these variational principles are
unrelated, and indeed one can prove that the variational principle of
\cite{BCDWKLM} can yield these other variational principles upon
solving a consistent subset of the extremality conditions. In some
cases, this obviously requires a proper extension of $\mathcal{V}$
with $\Upsilon$-dependendent terms and/or non-holomorphic corrections.

In the proposal of \cite{Ooguri:2004zv} the magnetic attractor
equations are imposed, so that the $Y^I$ are expressed in terms of the
magnetic charges $p^I$ and (real) electrostatic potentials $\phi^I$ at
the horizon,
\begin{equation}
\label{eq:electro-phi} 
  Y^I = \frac{\phi^I}{2\pi} + \frac{\mathrm{i} p^I}{2} \,.
\end{equation}
The electric attractor equations will then follow from a variational
principle associated with variations of the $\phi^I$ and the result
was written in the form,
\begin{eqnarray}
  \label{eq:real-F}
  \mathcal{S}_{\text{macro}}(p,q) &=& \mathcal{F}(\phi,p) - \phi^I\, 
  \frac{\partial\mathcal{F}(\phi,p)}{\partial \phi^I} \,,
 \nonumber\\ 
  q_I&=&   \frac{\partial\mathcal{F}(\phi,p)}{\partial \phi^I} \,,
\end{eqnarray}
where the real function $\mathcal{F}(\phi,p)$ is defined by 
 \begin{equation}
  \label{eq:def-real-F}
  \mathcal{F}(\phi,p) = 4\pi \,\mathrm{Im} [\,F(Y,\Upsilon)]_{\Upsilon=-64} \,.
\end{equation}
In \cite{CDWKM} it was shown how to incorporate non-holomorphic
corrections into the function ${\mathcal{F}}(\phi,p)$.

The result (\ref{eq:real-F}) shows that the function
$\mathcal{F}(\phi,p)$ and the entropy
$\mathcal{S}_{\text{macro}}(p,q)$ are related by a Legendre transform,
which suggests the introduction of a mixed black hole partition
function, $Z_{\text{BH}}(\phi,p)$, defined by
\begin{equation}
  \label{eq:mixed-partition}
\mathrm{e}^{\mathcal{F}(\phi,p)} = 
Z_{\text{BH}}(\phi,p)  = \sum_{\{q_I\}} \; d(q,p)\, 
\mathrm{e}^{ q_I \,\phi^I} \,,  
\end{equation}
where the $d(q,p)$ are the microscopic black hole degeneracies. This
partition function is a mixed partition function, as it treats the
electric and the magnetic charges differently: with respect to the
magnetic charges one is dealing with a microcanonical ensemble and
with respect to the electric charges one has a canonical ensemble.
Note that the left-hand side of (\ref{eq:mixed-partition}) can be
written as the modulus square of $\exp[-2\pi\mathrm{i}
F(Y,\Upsilon)]$, where the $Y^I$ are given by (\ref{eq:electro-phi})
and $\Upsilon=-64$. The holomorphic expression $\exp[-2\pi\mathrm{i}
F(Y,\Upsilon)]$ is actually related to the partition function for the
topological string; the non-holomorphic corrections, which we
suppressed here, are related to the so-called holomorphic anomaly
\cite{BCOV}. The connection with topological string theory was further
discussed in \cite{Verlinde:2004x,Ooguri:2005x}.

The equation (\ref{eq:mixed-partition}) implies that the black hole
degeneracies can be expressed as a Laplace transform of the partition
function $Z_{\text{BH}}(\phi,p)$ \cite{Ooguri:2004zv},
\begin{equation}
  \label{eq:laplace}
  d(q,p) \sim \int \prod_I \mathrm{d}\phi^I \, 
  \left\vert\mathrm{e}^{-2\pi\mathrm{i} F(Y,\Upsilon)}\right \vert^2 \; 
   \mathrm{e}^{ - q_I \,\phi^I} \,.  
\end{equation}
where the $Y^I$ are still given by (\ref{eq:electro-phi}) and
$\Upsilon=-64$.  For large values of the $q_I$ the Laplace transform
can be solved by a saddle-point approximation which leads to the
exponent of entropy $\mathcal{S}_{\text{macro}}(p,q)$ in accord with
(\ref{eq:real-F}).

One should be aware that there are, however, many subtleties with
these expressions (see, for instance, \cite{Dabholkar:2005x} where
some of these are discussed). One of them which we would like to
mention, is related to electric/magnetic duality. The attractor
equations (\ref{eq:attractor}) and the expression for the macroscopic
entropy (\ref{eq:W-entropy}) are manifestly consistent with this
duality. This means that, for instance, the expression for the entropy
transforms as a scalar function and its expression in a dual
description is simply obtained by applying the duality on the charges
$p^I$ and $q_I$, {\it i.e.},
$\mathcal{S}^\prime_{\text{macro}}(p^\prime,
q^\prime)=\mathcal{S}_{\text{macro}}(p,q)$. This implies, in
particular, that the entropy will be invariant under a subgroup of the
electric/magnetic duality group that constitutes an invariance. The
same property applies presumably also for the microscopic black hole
degeneracies, $d(q,p)$.\footnote{
  This is demonstrated for the case of heterotic black holes, where
  both the macroscopic and the microscopic description are invariant
  under $S$-duality. The application of more general dualities may be
  more subtle, however, as this particular example is outside the
  restricted context of the Wilsonian action.}
On the other hand, the mixed partition function and the functions
$F(Y,\Upsilon)$ and $\mathcal{F}(\phi,p)$ do not transform as
functions under electric/magnetic duality; this is already obvious
from the fact that the $(p^I,\phi^I)$ do not transform simply under
this duality, unlike $(p^I,q_I)$.  This aspect is, of course, relevant
when giving a more precise meaning to equations such as
(\ref{eq:mixed-partition}) and (\ref{eq:laplace}).

These subtle issues are, however, outside the scope of these lectures,
which are aimed at providing a pedagogical introduction and overview
of relatively recent results pertaining to the determination of the
black hole entropy in string theory and supergravity and their
relation. Interested readers are advised to consult the literature for
further information.

\vspace{4mm}
\noindent
Most of my own work on the topic of these lectures has been in
collaboration with Gabriel Lopes Cardoso, J\"urg K\"appeli and Thomas
Mohaupt, whom I thank for their valuable comments on the text. This
work is partly supported by EU contract MRTN-CT-2004-005104. 

\providecommand{\href}[2]{#2}
\begingroup\raggedright \endgroup

\end{document}